\documentclass[aps,prb,twocolumn,superscriptaddress,groupedaddress]{revtex4}
\usepackage{hyperref}
\usepackage{graphicx}
\usepackage{amsmath}
\usepackage{sidecap}
\usepackage{amsfonts}
\usepackage{xcolor}
\usepackage{color}
\usepackage{ulem}

\def\sro{Sr$_2$RuO$_4$}

\begin{document}
\title{$T_c$ and Pauli limited critical field of $\text{Sr}_2\text{Ru}\text{O}_4$: uniaxial strain
dependence}
\author{Yue Yu}
\affiliation{Department of Physics, Stanford University, 476 Lomita Mall, Stanford, CA 94305}
\author{Stuart Brown}
\affiliation{Department of Physics \& Astronomy, University of California, Los Angeles,  Los Angeles, CA
90095, USA}
\author{S. Raghu}
\affiliation{Department of Physics, Stanford University, 476 Lomita Mall, Stanford, CA 94305}
\affiliation{Stanford Institute for Materials and Energy Sciences, SLAC National Accelerator Laboratory,
Menlo Park, CA 94025, USA}
\author{Kun Yang}
\affiliation{National High Magnetic Field Laboratory and Department of Physics, Florida State
University, Tallahassee, Florida 32306, USA}

\begin{abstract}
Variations of critical temperature $T_c$ and in-plane critical field $H_{c2}$ of
$\text{Sr}_2\text{Ru}\text{O}_4$ under uniaxial stress have recently been reported. We compare the strain dependence of $T_c$ and $H_{c2}$
in various pairing channels ($d$-wave, extended s-wave and $p$-wave) with the experimental observations, by
studying a three-band tight-binding model that includes effects of spin-orbit and Zeeman couplings and
a separable pairing interaction. Our study helps narrow down the possibility of pairing channels. The importance of the multi-band nature of  $\text{Sr}_2\text{Ru}\text{O}_4$ is also highlighted.
\end{abstract}
\maketitle

\section{Introduction}
$\text{Sr}_2\text{Ru}\text{O}_4$ has long been one of the best characterized materials in which
unconventional superconductivity condenses out of a Fermi liquid~\cite{Mackenzie2003,Mackenzie2017}. Thus, it presents an almost unique
opportunity, where  well-controlled theoretical approaches can play a key role in deducing superconducting properties, starting from the underlying electronic structure. Nevertheless, several basic phenomenological
aspects, including the symmetry of the superconducting order parameter itself, remain unresolved. The results of early
NMR spectroscopy measurements \cite{Ishida1998} and spin-polarized neutron scattering studies \cite{Duffy2000},
together with evidence for time-reversal symmetry breaking (TRSB), \cite{Luke1998,Xia2006} were all
taken as consistent with a chiral $p_x+ip_y$ state. However, the chiral $p+ip$ state was recently excluded as a
possibility due to newly reported measurements of the $^{17}$O Knight shifts~\cite{Pustogow2019}, which
revealed a reduced spin susceptibility in the superconducting state. Moreover, the observation has been confirmed in independent NMR
studies \cite{Ishida2020}, as well as in a spin-polarized neutron scattering study~\cite{Petsch2020}.

The measurements reported in Ref. \onlinecite{Pustogow2019} are among several new experimental studies~\cite{Hicks2014,Steppke2017,Li2019,grinenko2020split}, in which the application of uniaxial ([100]) stress has placed further constraints on the nature of the \sro\ order parameter.   The induced strain in these experiments acts as a tetragonal symmetry breaking perturbation. Thus, it is a sensitive probe of multi-component order parameters that in turn are required for spontaneous TRSB in the superconducting state, and can be exploited to reveal more details of its nature.  
With these recent developments in mind, we are led to reconsider the phenomenological consequences and to see how distinct order parameters behave in the presence of strain.

The particular focus here is on recent experiments \cite{Steppke2017,Luo2019,Pustogow2019} of critical temperature
$T_c$ and in-plane critical field $H_{c2}$ in strained crystals.  We compute the strain
response of $T_c$ and $H_{c2}$ in different pairing channels and compare them with the observations. At the so-called Van Hove strain ($\varepsilon_{aa}=\varepsilon_v$), one of the Fermi sheets, customarily labeled  $\gamma$ in the literature, crosses the
Van Hove singularity (vHs) at the boundary of the first Brillouin zone. This Fermi sheet consists of quasiparticles built predominantly from electrons in the $d_{xy}$ orbital, with weak mixing of $d_{xz}, d_{yz}$ orbitals in the presence of atomic spin-orbit coupling.  Since the $\gamma$ sheet has little dispersion in the $c$-direction, the density of states is expected to
diverge logarithmically in the neighborhood of the vHs. Therefore, tuning $E_F$ to the vHs results in an expected enhancement of both the transition temperature \cite{Steppke2017} and the upper
critical field~\cite{Luo2019}. Further, the enhancement of in-plane ($\vec{H}//\mathbf{b}$) critical field was observed to be stronger than that in the critical temperature \cite{Steppke2017,Luo2019}. Here, we compare and contrast the observed behavior to expectations for selected  order parameter symmetries.

More specifically, in this work, we analyze the ratio $H_{c2}/T_c$ as a function of uniaxial $\varepsilon_{xx}-\varepsilon_{yy}$ strain, by studying BCS theory on a 3-band tight binding model for different pairing channels,
including $d$-wave, $p$-wave and extended-s-wave pairing channels, and compare the results with experimental
observations. Our study points out a new direction for narrowing down the possible choices of order
parameters for  $\text{Sr}_2\text{Ru}\text{O}_4$, and the methods are readily applied to other systems.
Besides comparing results obtained for different pairing channels, the importance of the multi-band
nature of this material is highlighted by comparing results with/without atomic spin-orbit coupling (SOC) and
orbital Zeeman effects. Guided by the observation of a field-induced first order transition from the (low-field) superconducting state \cite{Yonezawa2014}, we consider the possibility of an inhomogeneous Fulde-Ferrell-Larkin-Ovchinnikov (FFLO) state \cite{Mackenzie2017}. Our study of the strain-dependent $H_{c2}/T_c$ in \sro\ may provide a new way to search for the FFLO
state elsewhere.

As we explain in more detail below, our key finding is that (1) the strain-dependence of the ratio
$H_{c2}/T_c$ in d+extended-s-wave pairing channel is consistent with the experimental observations,
while $p$-wave pairing channel (analogs of B-phase of ${}^3\text{He}$) is not. (2) The multi-band nature of
$\text{Sr}_2\text{Ru}\text{O}_4$, including spin-orbit coupling and the orbital Zeeman effect is necessary
for the correct dependence. (3) FFLO pairing of quasiparticles on the $d_{xy}$ orbital is not sensitive to the Van Hove strain, due to
Fermi surface nesting.

The paper is organized as follows. In Sec.\ref{model}, we introduce the settings in BCS theory and the
band structure. In Sec. \ref{results}, we present the numerical results for different pairing channels
with/without multi-band effects. In Sec.\ref{FFLO}, we extend our study to the FFLO state.

\section{The model}
\label{model}
We consider an effective three-band tight-binding-Hamiltonian  for $t_{2g}$  ($d_{yz}$,$d_{xz}$,$d_{xy}$)
electrons of $\text{Sr}_2\text{Ru}\text{O}_4$ with tetragonal symmetry under Zeeman effect. The
Hamiltonian is given by $H_0+H_{Z}+H_{BCS}$, where:
\begin{equation}
H_0=\sum_{\vec{k},a,b,\sigma}h_0^{ab}(\vec{k})c_{\vec{k}a\sigma}^{\dagger}c_{\vec{k}b\sigma}+H_{SOC}
\end{equation}
, and:
\begin{equation}
\begin{split}
&h_0(\vec{k})=\left[\begin{array}{ccc}
\epsilon^{yz}&\epsilon^{off}&\\
\epsilon^{off}&\epsilon^{xz}&\\
&&\epsilon^{xy}
\end{array}\right]\\
&\epsilon^{yz}=-2t_2\tau\cos{k_x}-2t_1/\tau\cos{k_y}-\mu\\
&\epsilon^{xz}=-2t_1\tau\cos{k_x}-2t_2/\tau\cos{k_y}-\mu\\
&\epsilon^{xy}=-2t_3(\tau\cos{k_x}+1/\tau\cos{k_y})\\
&-4t_4\cos{k_x}\cos{k_y}-2t_5\cos(2k_x)\cos(2k_y)-\mu\\
&\epsilon^{off}=-4t_6\sin{k_x}\sin{k_y}.
\end{split}
\label{e1}
\end{equation}
Here, $c_{\vec{k}a\sigma}^\dag \left(c_{\vec{k}a\sigma} \right)$ are creation(annihilation) operators
for electrons in $a=d_{yz}$, $d_{xz}$, or $d_{xy}$ orbitals for spin state $\sigma = \uparrow, \downarrow$, and $h_0$ is a $3\times{3}$ Hamiltonian in orbital space. The
parameters are obtained in Ref. \cite{Zabolotnyy2013} by fitting the above TBH with experimental data, and the resulted fitting parameters are listed here
$(t_1,t_2,t_3,t_4,t_5,t_6,\mu)=(0.145,0.016,0.081,0.039,0.005,0,0.122)eV$. Note that the off-diagonal term $\epsilon^{off}$ that couples $d_{yz}$ and $d_{xz}$ orbitals is zero from fitting. Here, the three $t_{2g}$
orbitals $(d_{yz}$, $d_{xz}$,$d_{xy})$ transform as a vector under point-group symmetry operations. Hence, the
angular momentum operator in this internal coordinate representation is $\mathcal L^{a}_{bc} = -i
\epsilon_{abc}$, where $\epsilon_{abc}$ is the totally anti-symmetric tensor, while spin operators are the standard Pauli matrices. Thus, the spin-orbit coupling is
\begin{equation}
H_{SOC}=\vec{L}\cdot{\vec{S}}=\lambda
\left(\begin{array}{ccc}
0& i\sigma^z&-i\sigma^y \\
-i\sigma^z&0& i\sigma^x\\
 i\sigma^y&-i\sigma^x&0.
\end{array}\right)
\end{equation}
The strength of spin-orbit coupling is taken to be $\lambda=0.032eV$. \cite{Zabolotnyy2013}
\begin{figure}[h]
\centering
\includegraphics[scale=0.23]{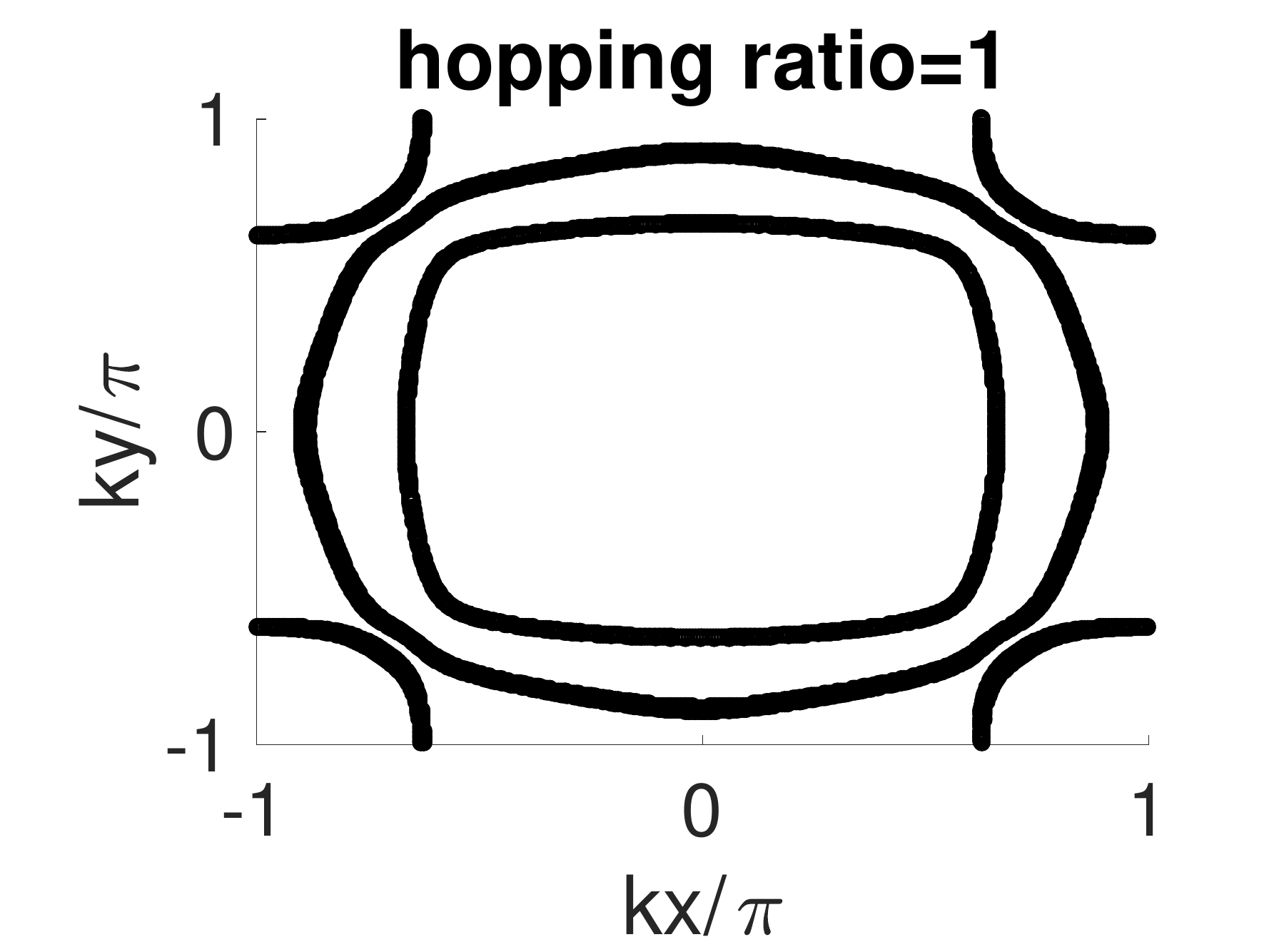}
\includegraphics[scale=0.23]{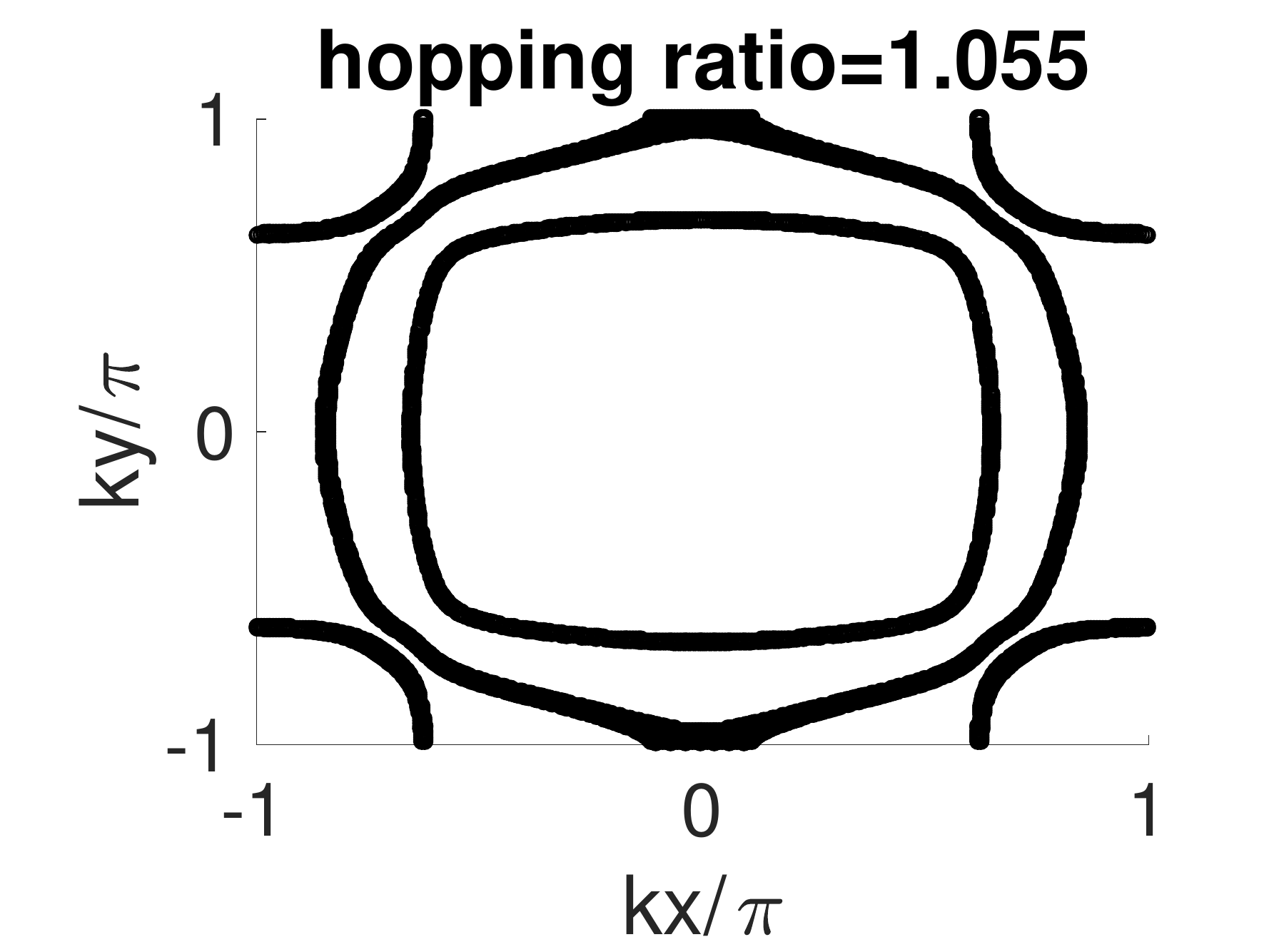}
\caption{Fermi surfaces of the three band tight binding Hamiltonian in Eq.\ref{e1} (Left)Zero strain
$\tau=1$, where system has tetragonal symmetry. (Right) Van Hove strain around $\tau=1.055$, where the
$\gamma$ band touches the boundary of the first Brillouin zone. }
\label{f1}
\end{figure}

We introduce the ``hopping ratio" $\tau$ to incorporate the effect of uniaxial strain, which modifies
the hopping strength along x and y direction (nearest neighbor hopping strength $t_1$ and $t_2$) in Eq.~\ref{e1}.
Under the above settings, zero uniaxial strain corresponds to hopping ratio $\tau=1$, while Van Hove
strain is around $\tau=1.055$. Fermi surfaces at zero strain and Van Hove strain are plotted in
Fig.\ref{f1} .

The Zeeman field couples to both the spin and  orbital, and the resulting Zeeman term is
\begin{equation}
H_Z=-\vec{H}\cdot{\vec{\sigma}}\otimes{\tau_0}+\sigma^0\otimes
\left(\begin{array}{ccc}
0& iH_z&-iH_y \\
-iH_z&0&iH_x\\
iH_y&-iH_x&0
\end{array}\right).
\end{equation}
Here, we have assumed that the system is strongly type-II, so that $\vec{H}$ is the external magnetic
field. $\tau_0$ is the identity matrix in orbital space, while $\sigma^0$ is the identity matrix in spin
space.

In this work, we will consider an approximation, that the order parameters are purely on the $d_{xy}$ orbital
of $\text{Sr}_2\text{Ru}\text{O}_4$, which are closest to tbe $\gamma$ band and most sensitive to the Van Hove
strain.  It should be noted that order parameters on other bands also contribute to the total gap
function. However, they are much less strain-sensitive, so variations of $T_c$, $H_{c2}$ due to pairings on the other bands would otherwise be smooth and analytic. Experimentally, which band
contributes mostly to superconductivity is still being investigated~\cite{Mackenzie2017}.
The general form of the BCS interaction on the $d_{xy}$ orbital can be written as
\begin{equation}
\begin{split}
H_{\text{BCS}}=-\sum_{\vec{k},\vec{k}',\lbrace{\sigma_i}\rbrace}V_{\sigma_1\sigma_2\sigma_3\sigma_4}(\vec{k},\vec{k}'){c}_{-\vec{k}\sigma_1}{c}_{\vec{k}\sigma_2}{c}_{\vec{k}'\sigma_3}^{\dagger}c_{-\vec{k}'\sigma_4}^{\dagger}
\end{split}
\end{equation}
Here, $\sigma_i$ denotes spin up/down. In the following calculations, we assume for simplicity that the above BCS
interaction is separable, i.e. it is of the following form,
\begin{equation}
\begin{split}
&V_{\sigma_1\sigma_2\sigma_3\sigma_4}(\vec{k},\vec{k}')=g\,
f(\vec{k})^\dagger_{\sigma_2\sigma_1}f(\vec{k'})_{\sigma_3\sigma_4}\\
&\hat{\Delta}_{\sigma_1\sigma_2}(\vec{k})=\Delta\,{f}(\vec{k})_{\sigma_1\sigma_2}
\end{split}\label{e5}
\end{equation}
Then the BCS gap equation can be simplified to
\begin{equation}
\begin{split}
\Delta=\sum_{|\tilde{\epsilon_k}|<E_{D},\sigma_1\sigma_2}gf(\vec{k})_{\sigma_1\sigma_2}\langle{c}_{-\vec{k}\sigma_1}{c}_{\vec{k}\sigma_2}\rangle.
\end{split}\label{e6}
\end{equation}
Here $\langle...\rangle$ denotes thermal averaging, and $\tilde{\epsilon_k}$ is the normal state
eigenenergy.
$E_{D}$ is an energy cutoff, analogous to the Debye temperature in conventional BCS theory. Given a pairing
channel $f(\vec{k})$ and pairing strength $g$, we can numerically diagonalize the Bogoliubov–de Gennes (BdG) Hamiltonian and
solve for the gap magnitude $\Delta$ self-consistently, at arbitrary temperature $T$, magnetic field $H$ and
hopping ratio $\tau$. The critical temperature can then be determined by the standard
procedure. When calculating the response to an in-plane critical field, we neglect the $c$-axis warping of the Fermi surface, and consider a
2-dimensional Fermi surface. We thus neglect the orbital effect of the in-plane magnetic field, and
the resulting $H_{c2}$ is the Pauli-limited critical field.

In order to better compare the results from different pairing channels, we would like to the fix the gap
magnitude at zero temperature, zero magnetic field and zero uniaxial strain to be the same in all
channels, i.e.
\begin{equation}
\Delta(T=0,B=0,\tau=1)=\Delta_0\equiv2.8\times{10^{-4}}eV.
\label{e7}
\end{equation}
The magnitude of $\Delta_0$ is chosen, such that $T_c$ is $\text{O}(1)$ Kelvin, which is
on the same order as the experimental value \cite{Steppke2017}. The above fixing is achieved by tuning
the BCS interaction strength $g$ in each channel. Those interaction strengths will then be fixed
throughout the calculation. That is, we have assumed that the BCS interaction strength g is
strain-independent. A strain-dependent interaction strength will change the magnitude of $T_c$ and
$H_{c2}$. However, the interaction strength will not affect the ratio $T_c/H_{c2}$. This is well-known
in the standard BCS theory without disorder, where $T_c$ and Pauli-limited $H_{c2}$ are both proportional to the gap
magnitude at zero temperature and zero field, while the proportionality constant only depends on the band
structure and type of pairing channel, rather than on the interaction strength.  \cite{chandrasekhar1962}
The energy cutoff of the interaction (analogous to Debye temperature for BCS theory) is taken to be
$E_D=10\Delta_0$.

\section{results}
\label{results}
In the first subsection below, we present numerical results for $d$-wave, $p$-wave and
$d$+$s$-wave pairing channels, in the absence of spin-orbit coupling and the orbital Zeeman effect.
In the next subsection, we will perform the same calculations, but with SOC and
orbital Zeeman effects.

 The direct comparison between various order parameters helps narrow down the possible pairing
 channels in $\text{Sr}_2\text{Ru}\text{O}_4$.
We will also highlight the importance of the multi-band nature of $\text{Sr}_2\text{Ru}\text{O}_4$, as
we compare the results between these two subsections. It is worth noting that, in the absence of SOC and
orbital Zeeman effect, the problem becomes effectively single-orbital.

\subsection{single-band}
In this subsection, we remove the spin-orbit coupling and orbital Zeeman effect in the Hamiltonian. Now
the $d_{yz}$ and $d_{xz}$ orbitals will not affect the calculations, and effectively we end up with the
problem on the $d_{xy}$ orbital.
It should be noted that the Van Hove strain is shifted to a larger hopping ratio, $\tau=1.08$.

\subsubsection{$d$-wave pairing channel}
\begin{figure}[h]
	\centering
	\includegraphics[width=4.2cm]{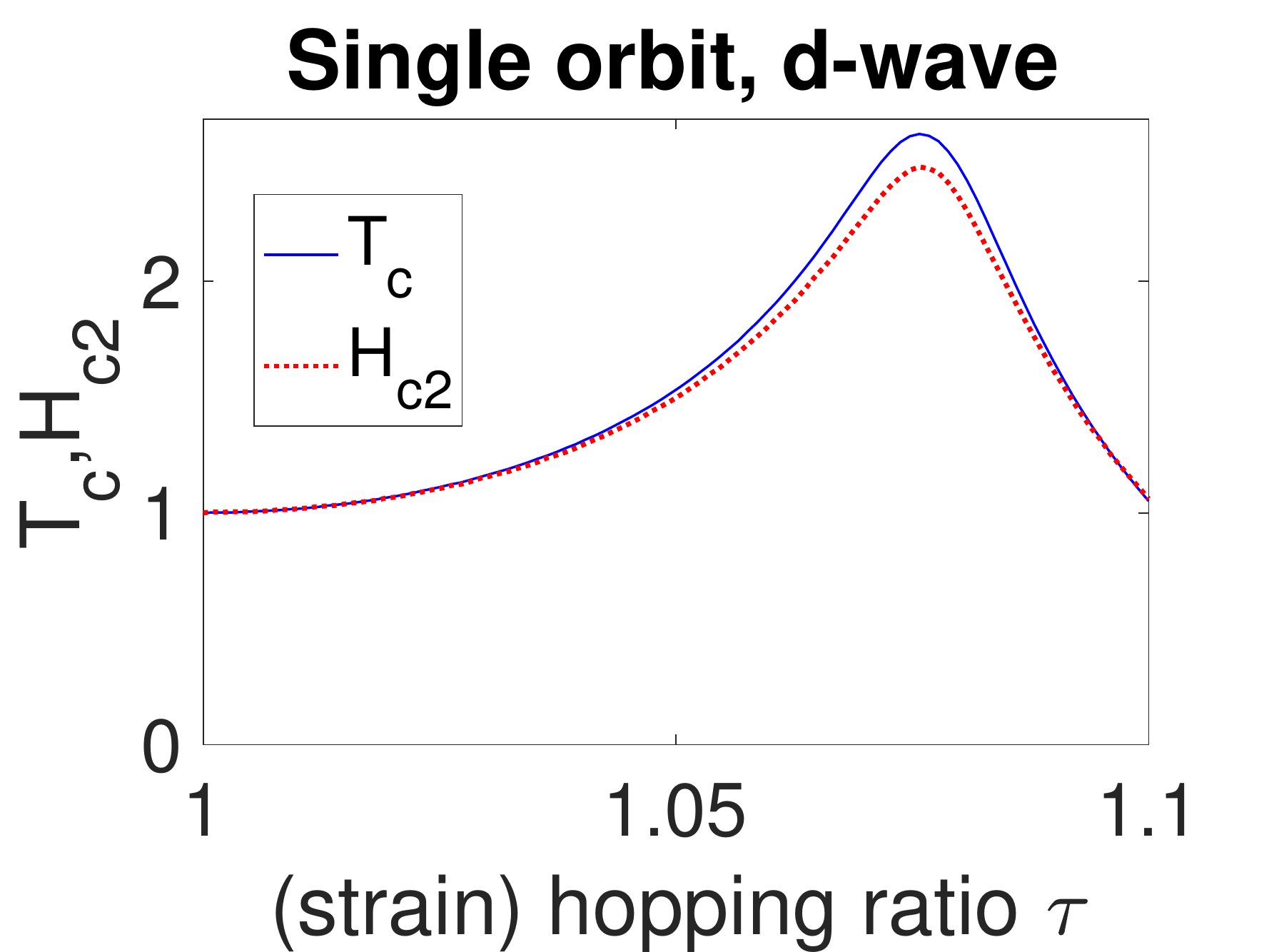}
	\includegraphics[width=4.2cm]{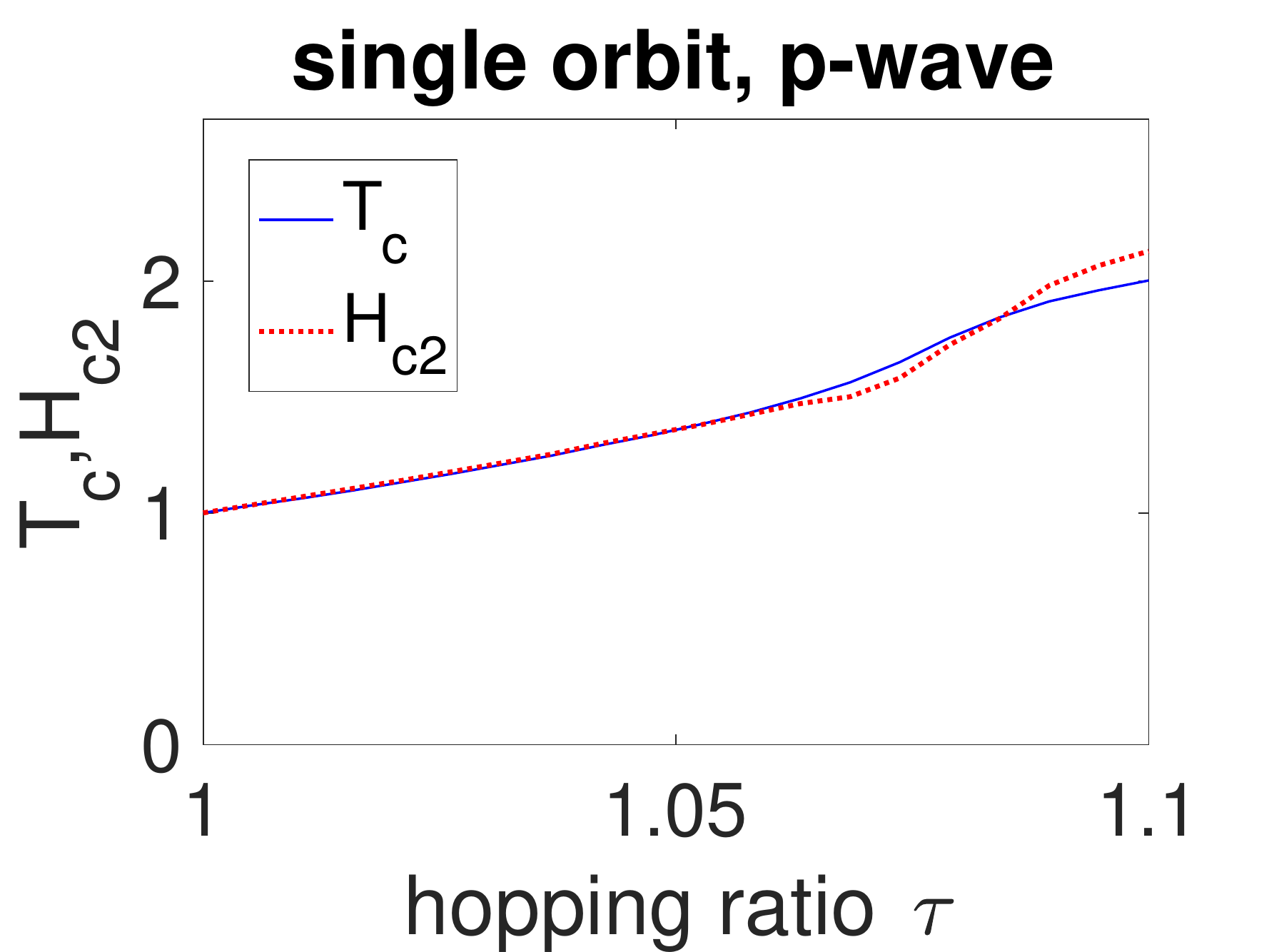}
	\caption{Critical temperature $T_c$ and Pauli-limited critical field $H_c$ for $d$-wave and $p$-wave
pairing on single-orbital model, as a function of uniaxial strain (hopping ratio $\tau$). Both quantities
have been normalized to unity at zero strain ($\tau=1$). (Left) For $d$-wave pairing, the peak position for
the two quantities is at the Van Hove strain. $H_{c2}$ clearly has a weaker enhancement than $T_c$ when
approaching the Van Hove strain, which is inconsistent with experimental observation.
		(Right) For $p$-wave pairing, $H_{c2}$ and $T_c$ are insensitive to the Van Hove singularity at
$\tau=1.077$, since we have assumed that the BCS interaction is strain-independent and the $p$-wave gap
function Eq.\ref{e10} vanishes at
$(k_x,k_y)=(0,\pi)$. The ratio $H_{c2}/T_c$ decreases when approaching the Van Hove strain, which is
inconsistent with experimental observation.}
	\label{f3}
\end{figure}

Let us start with the $d$-wave pairing channel 
\begin{equation}
{f}({\vec k})_{\sigma_1 \sigma_2}=i\sigma^y_{\sigma_1 \sigma_2} (\cos{k_x}-\cos{k_y}).
\end{equation}
  $T_c$ and Pauli-limited critical field $H_{c2}$ as a function of uniaxial strain (hopping
ratio $\tau$) are shown in the left panel of Fig.~\ref{f3}. Both quantities have been normalized to
unity at zero strain ($\tau=1$).

At the Van Hove strain $\tau=1.077$, $T_c$ (blue solid line) is clearly enhanced more significantly than
$H_{c2}$ (red dotted line). Therefore, the ratio $H_{c2}/T_c$ decreases as approaching the Van Hove
strain, which is inconsistent with the experimental observations.

\subsubsection{$p$-wave pairing channel}
\label{Secp1}
For spin-triplet superconductors without spin-orbit coupling, Pauli-limited critical field cannot be
obtained, if the $\hat{d}$-vector of the pairing state is perpendicular to the magnetic field. With the presence
of SOC (in next subsection), the above scenario no longer holds, but the resulting $H_{c2}$ could be
much bigger than the maximal gap magnitude if SOC is small. Experimentally~\cite{Luo2019}, $H_{c2}$ is found to be of the same order of the maximal gap magnitude, $\Delta/g\mu_B$. Noting also that the strain lifts the $p_x$, $p_y$ degeneracy, states with $d$-vector parallel to the magnetic field are, in principle, possible.
That is, the $p$-wave pairing channel of the form:
\begin{equation}
{f}({\vec k})_{\sigma_1 \sigma_2}=i(\sigma^x\sigma^y)_{\sigma_1 \sigma_2}\sin{k_x}
\label{e10}
\end{equation}
, arising on the $d_{xy}$ orbital, is considered. The $\hat{d}$-vector is along $x$-axis, and we calculate the corresponding $H_{c2}$.

Critical temperature $T_c$ and Pauli-limited critical field $H_{c2}$ as a function of uniaxial strain
(hopping ratio $\tau$) are shown in the right panel of Fig. \ref{f3}. Both quantities are normalized to unity at zero strain ($\tau=1$).
$T_c$ (blue solid line) is clearly enhanced more significantly than $H_{c2}$ (red dotted line) at the
Van Hove strain $\tau=1.077$. Therefore, the ratio $H_{c2}/T_c$ decreases as approaching the Van Hove
strain, which is inconsistent with the experimental observations.

Under the assumption of strain-independent BCS interaction strength, the critical temperature and
critical field in the $p$-wave pairing channel are not sensitive towards the Van Hove strain, and we do
not observe any peak in Fig. \ref{f3} at the Van Hove strain. This is because the $p$-wave gap function $f({\vec k})$ vanishes at the Van Hove singularity $(k_x,k_y)=(0,\pi)$.

\subsubsection{$d$+extended-$s$-wave pairing channel}\label{Secds1}
We now consider mixture between $d$-wave ${f}({\vec k})_{\sigma_1 \sigma_2}=i\sigma^y_{\sigma_1 \sigma_2} (\cos{k_x}-\cos{k_y})$ and extended-$s$-wave
${f}({\vec k})_{\sigma_1 \sigma_2}=i\sigma^y_{\sigma_1 \sigma_2} (\cos{k_x}+\cos{k_y})$ pairing channel. When applying uniaxial strain, the
tetragonal symmetry is broken, and these two pairing channels belong to the same irreducible representation, and hence are allowed to mix.
In the following calculations, we again only consider pairing on the $d_{xy}$ orbital. We now introduce two
BCS interaction strengths $g_d$ and $g_s$ for the two channels, and assume they are strain-independent.
We choose the strengths $g_d$ and $g_s$, such that the $d$-wave gap magnitude satisfies Eq.\ref{e7}, and
the gap magnitude of extended-$s$-wave pairing channel vanishes at zero strain.

\begin{figure}[h]
 	\centering
 	\includegraphics[width=4cm]{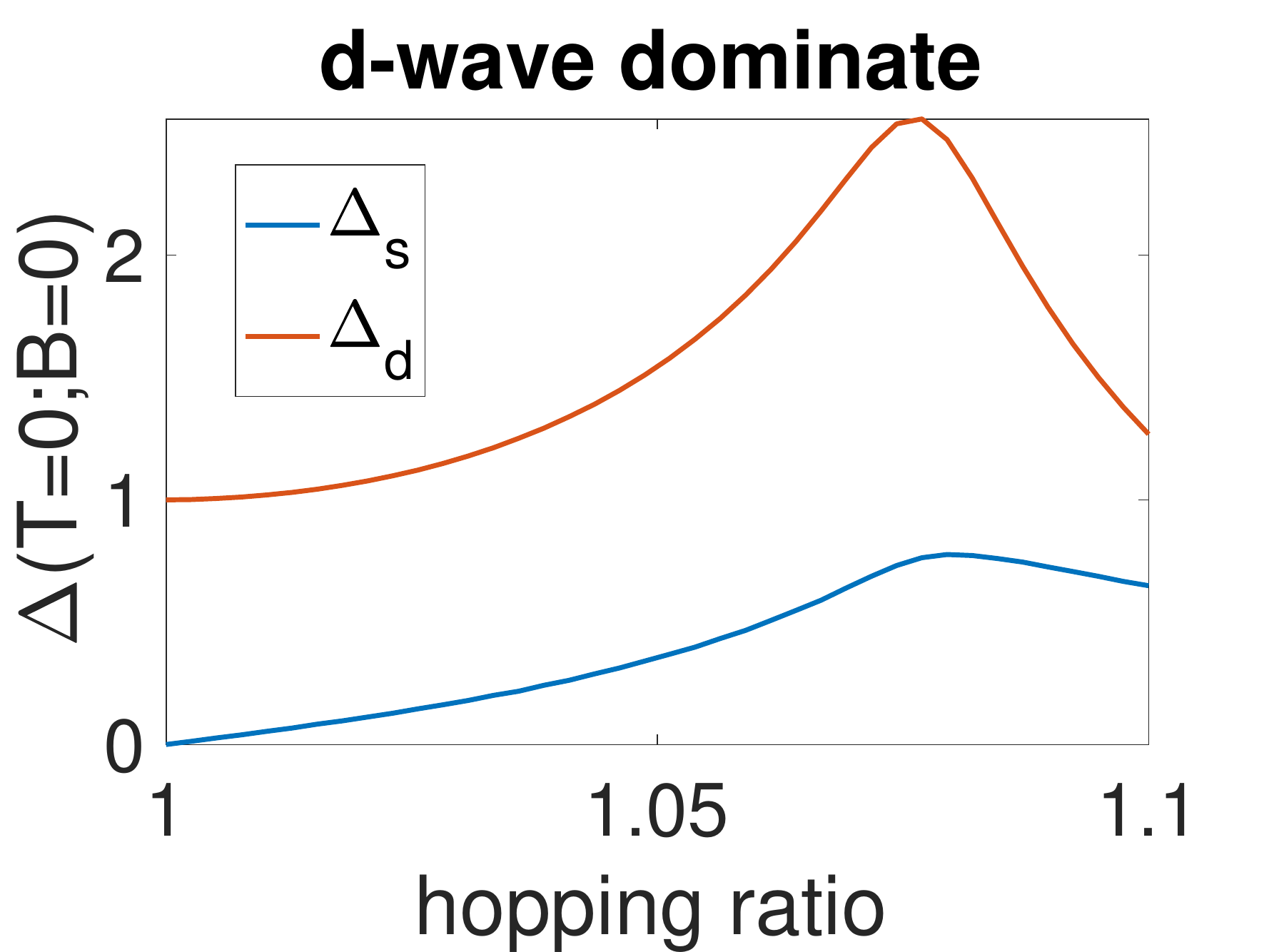}
 	\includegraphics[width=4cm]{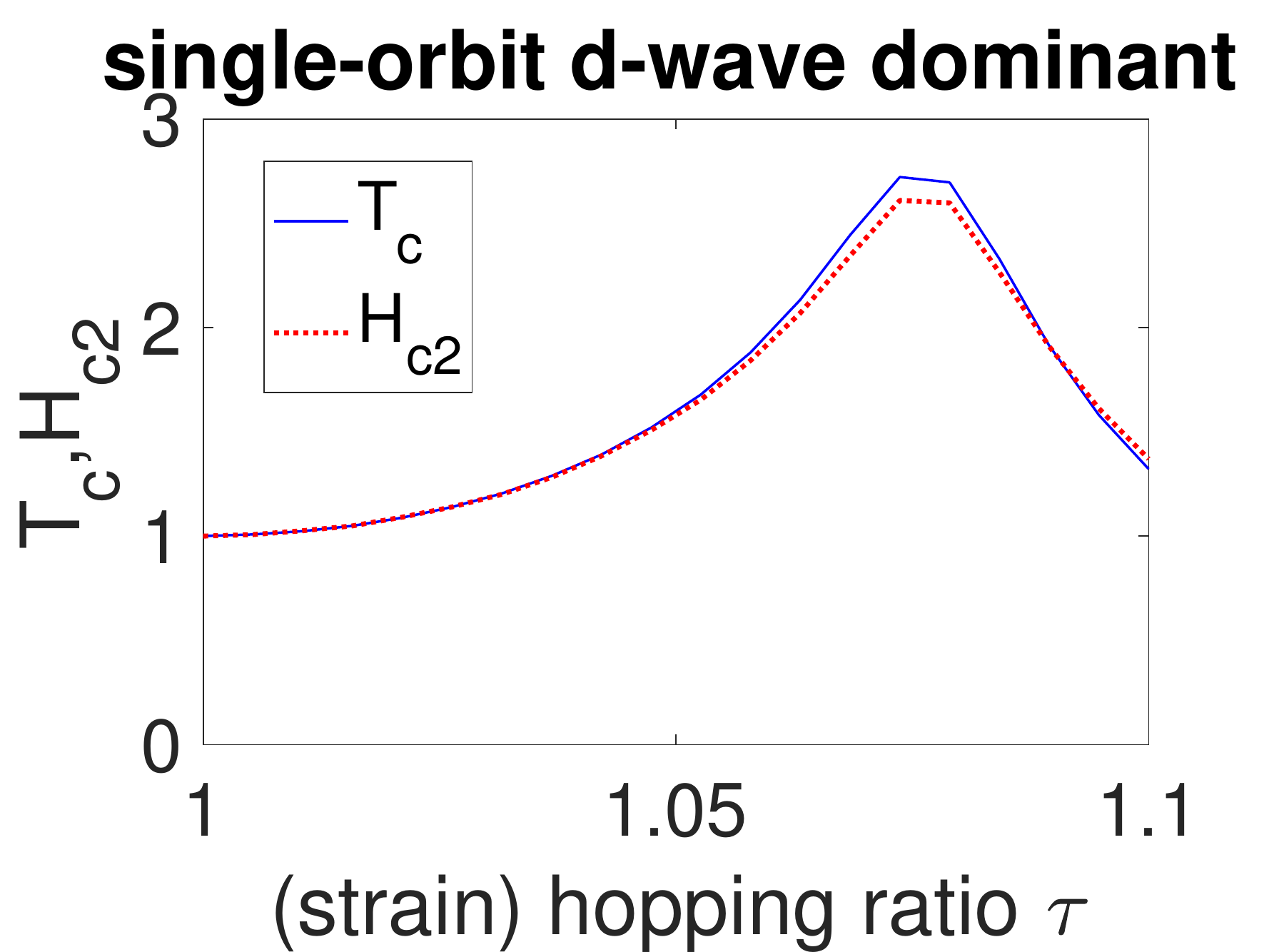}
 	\caption{Results for mixture of extended-$s$-wave and $d$-wave pairing channels on a single-orbital model. A
 particular set of interaction strength $(g_d,g_s)$ has been chosen. (Left) Pairing magnitude at $T=H=0$
 for each pairing channel as a function of strain.  (Right) critical temperature and critical field, as
 a function of strain. The enhancement in $H_{c2}$ is weaker than that in $T_c$, which is inconsistent
 with the experimental observation.}
 	\label{f5}
 \end{figure}

 The gap magnitude as a function of uniaxial strain is plotted in the left panel of Fig.\ref{f5}. Thus, in
 this calculation, $d$-wave pairing dominates over the extended-$s$-wave pairing. We solved the two gap
 equations, and obtained critical temperature and critical field.

 At the Van Hove strain $\tau=1.077$, $T_c$ (blue solid line) is clearly enhanced more significantly
 than $H_{c2}$ (red dotted line). Therefore, the ratio $H_{c2}/T_c$ decreases on approaching the Van Hove strain, which is inconsistent with the experimental observations.

In this subsection, we effectively removed $d_{yz}$ and $d_{xz}$ orbitals from the Hamiltonian, and
obtained $H_{c2}$ and $T_c$ for a single orbital ($d_{xy}$ orbit) system. We have tried $d$-wave, $p$-wave and
$d$+extended-$s$-wave pairing channels, but none trend similarly to the experimentally observed ratio $H_{c2}/T_c$.

\subsection{three-band}
\subsubsection{$d$-wave pairing channel}
\label{Secd3}
\begin{figure}[h]
\centering
\includegraphics[width=4.2cm]{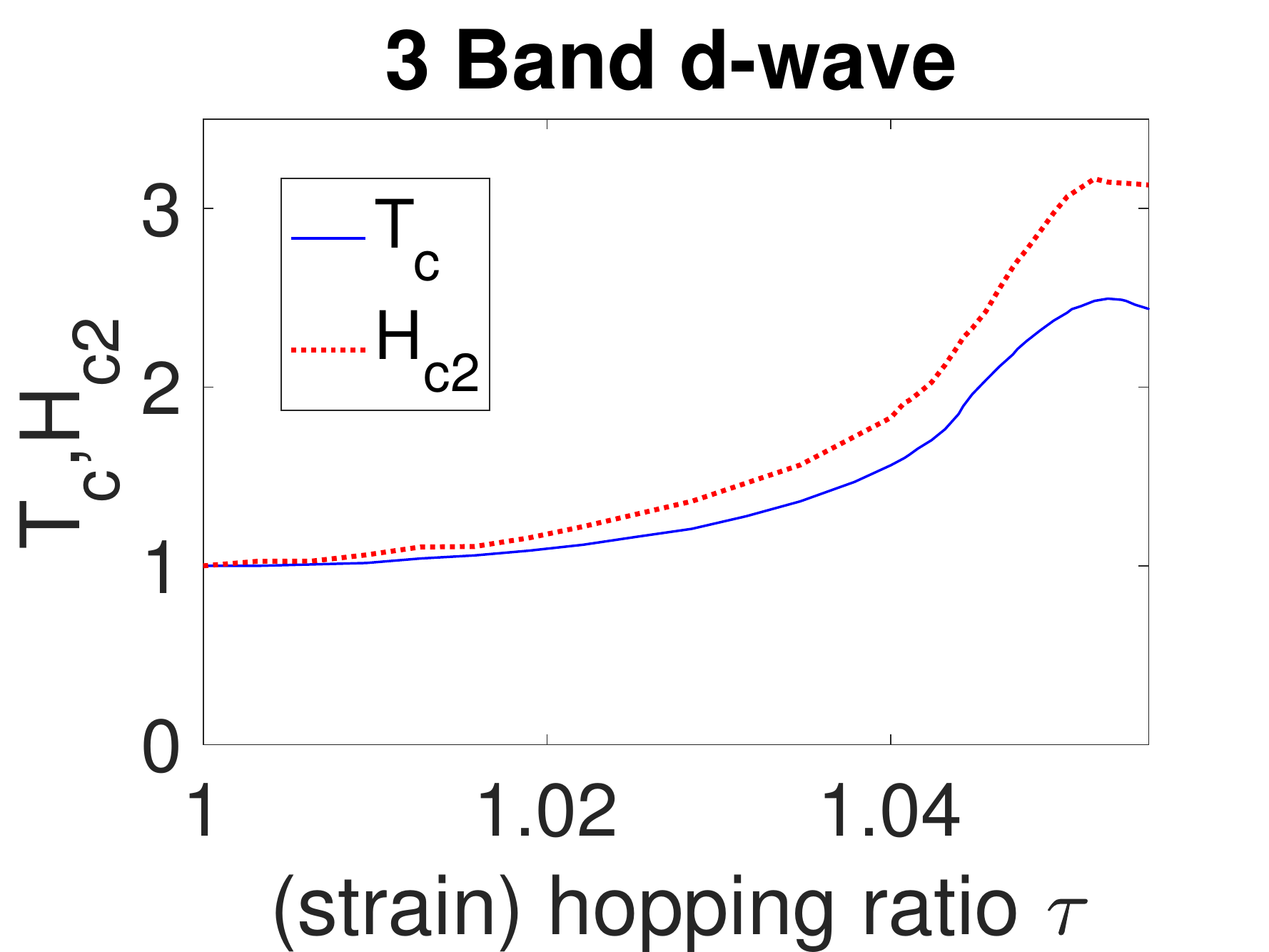}
\includegraphics[width=4.2cm]{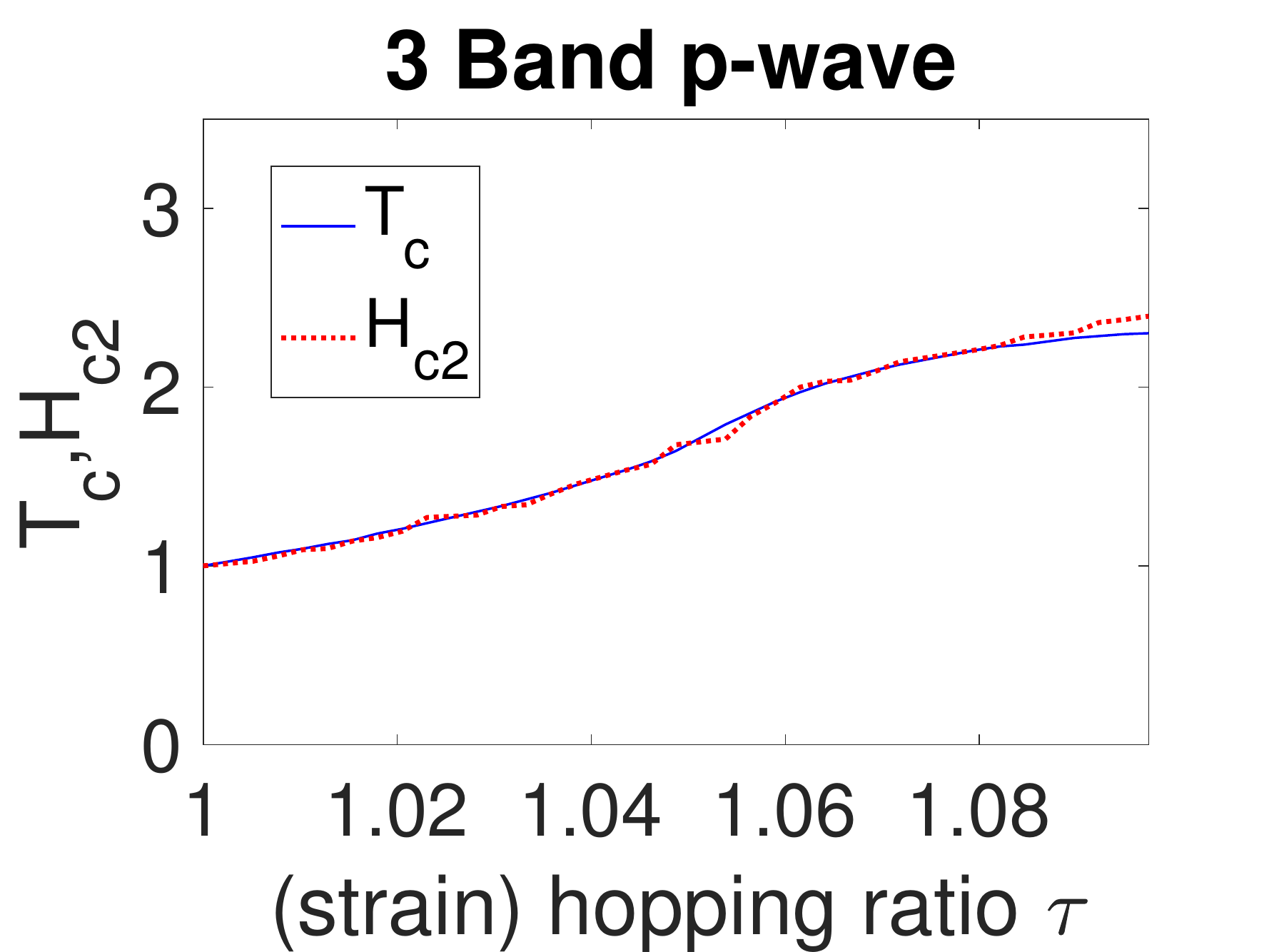}
\caption{Critical temperature $T_c$ and Pauli-limited critical field $H_c$ for $d$-wave and $p$-wave pairing
on a 3-band model, as a function of uniaxial strain (hopping ratio $\tau$). Both quantities have been
normalized to unity at zero strain ($\tau=1$). (Left) For $d$-wave pairing, the peak position for the two
quantities is at the Van Hove strain. $H_{c2}$ clearly has a stronger enhancement than $T_c$ when
approaching the Van Hove strain, which is consistent with experimental observation.
(Right) For $p$-wave pairing, $H_{c2}$ and $T_c$ are not sensitive to the Van Hove singularity, since we
have assumed that the BCS interaction is strain-independent and the $p$-wave gap function vanishes at the Van Hove singularity $(k_x,k_y)=(0,\pi)$. The
ratio $H_{c2}/T_c$ does not change when approaching the Van Hove strain, which is inconsistent with
experimental observation.}
\label{f2}
\end{figure}

We start with $d$-wave only pairing channel ${f}({\vec k})_{\sigma_1 \sigma_2}=i\sigma^y_{\sigma_1 \sigma_2} (\cos{k_x}-\cos{k_y})$ for the 3-band system.
The critical temperature $T_c$ and Pauli-limited critical field $H_{c2}$ as a function of uniaxial strain
(hopping ratio $\tau$) are shown in the left panel in Fig. \ref{f2}. Both quantities have been
normalized to unity at zero strain (at $\tau=1$).

$H_{c2}$ (red dotted line) is clearly enhanced more significantly than $T_c$ (blue solid line).
Therefore, the ratio $H_{c2}/T_c$ increases as approaching the Van Hove strain, which is consistent with
the experimental observations.

\subsubsection{$p$-wave pairing channel}
For reasons mentioned in Sec.\ref{Secp1}, in order to calculate the Pauli-limited critical field, the $p$-wave pairing state with ${f}({\vec k})_{\sigma_1 \sigma_2}=i(\sigma^x\sigma^y)_{\sigma_1 \sigma_2}\sin{k_x}$ is considered. The $\hat{d}$-vector
is along $x$ direction, and we calculate critical field also in this direction.

$T_c$ and Pauli-limited critical field $H_{c2}$ as a function of uniaxial strain
(hopping ratio $\tau$) are shown in the right panel of Fig. \ref{f2}. Both quantities have been
normalized to unity at zero strain (at $\tau=1$).
The enhancement in $H_{c2}$ (red dotted line) and $T_c$ (blue solid line) are almost the same.
Therefore, the ratio $H_{c2}/T_c$ does not change on approach to the Van Hove strain, which is
inconsistent with the experimental observations.

For the same reasons as  in Sec.\ref{Secp1}, we did not observe any peak in $T_c$ or $H_{c2}$ near the
Van Hove strain, since the gap function for the $p$-wave pairing state vanishes at the Van Hove singularity.
Again, one could get the correct shape of peak by introducing strain-dependent BCS interaction strengths,
but this will not affect the ratio $H_{c2}/T_c$.

\subsubsection{d+extended-s-wave pairing channel}
\begin{figure}[h]
	\centering
	\includegraphics[width=4cm]{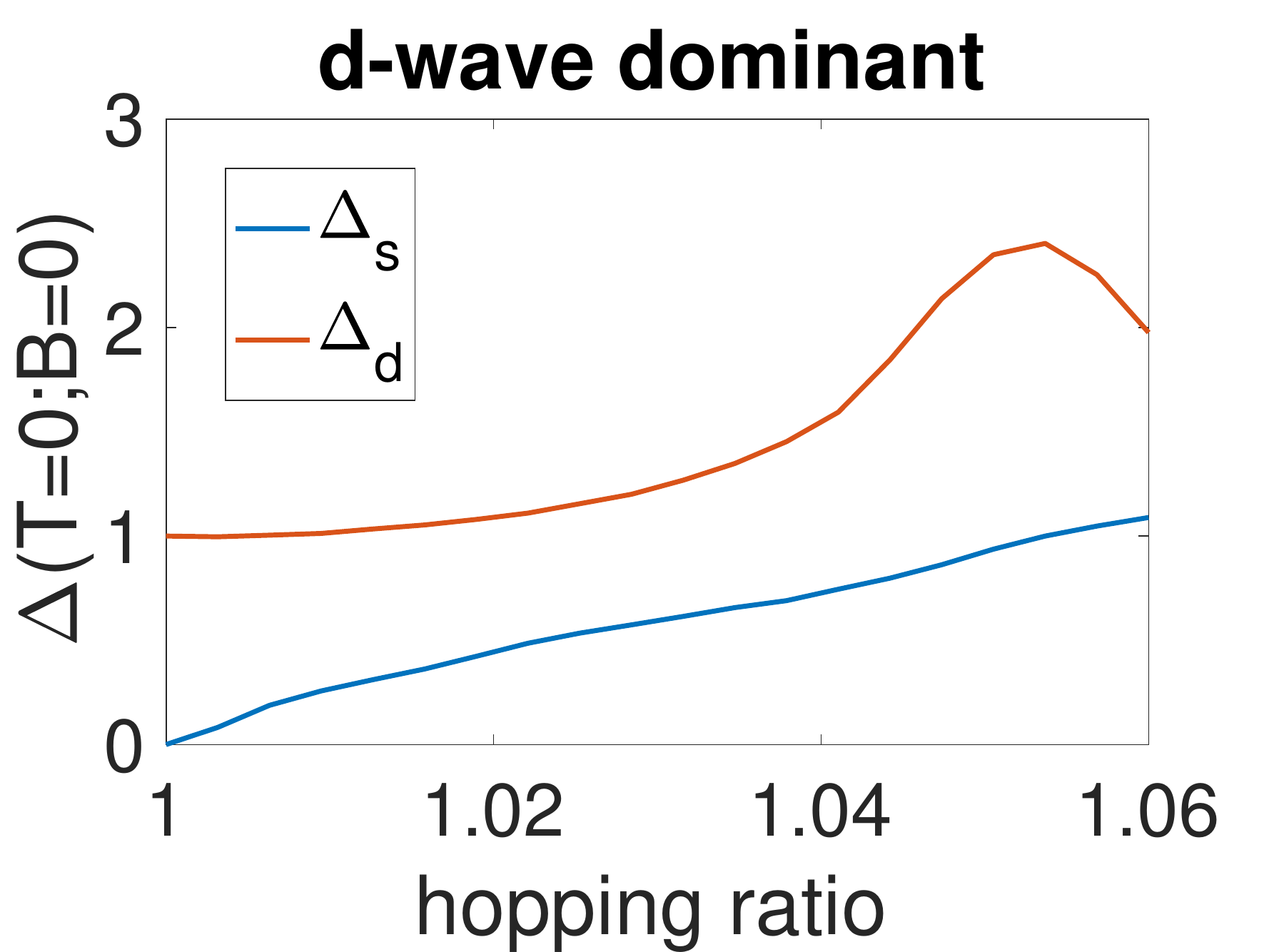}
	\includegraphics[width=4cm]{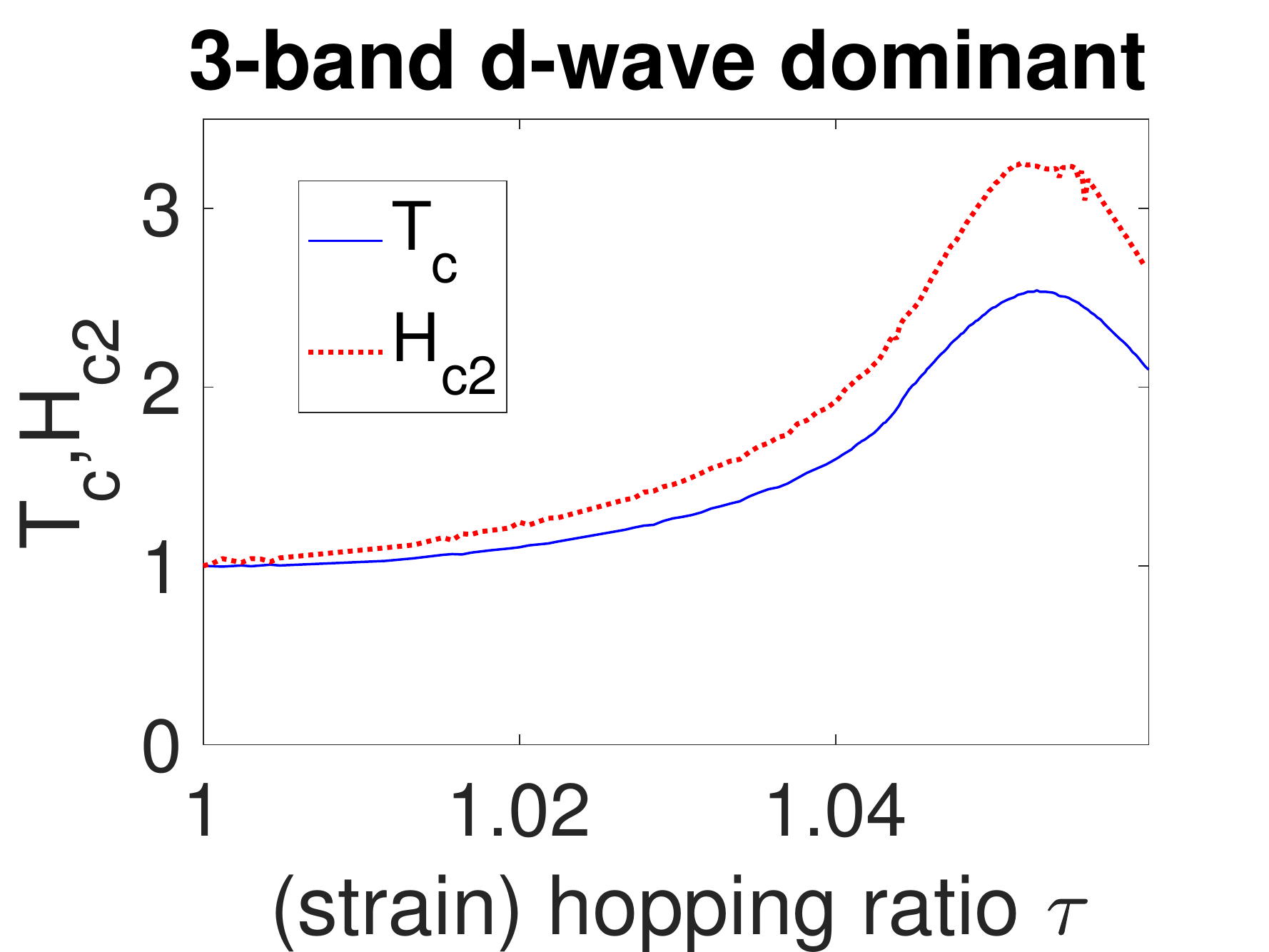}
	\caption{Results for mixture of extended-$s$-wave and $d$-wave pairing channels on 3-band model. A
particular set of interaction strengths $(g_d,g_s)$ has been chosen. (Left) Pairing magnitude at $T=H=0$
for each pairing channel as a function of strain.  (Right) critical temperature and critical field, as a
function of strain. The enhancement in $H_{c2}$ is stronger than that in $T_c$, which is consistent with
the experimental observation.}
	\label{f4}
\end{figure}

We now turn to mixture between $d$-wave ${f}({\vec k})_{\sigma_1 \sigma_2}=i\sigma^y_{\sigma_1 \sigma_2} (\cos{k_x}-\cos{k_y})$ and extended-$s$-wave
${f}({\vec k})_{\sigma_1 \sigma_2}=i\sigma^y_{\sigma_1 \sigma_2} (\cos{k_x}+\cos{k_y})$ pairing channel. Similar to the single-orbital case in Sec.
\ref{Secds1} , we choose the strength $g_d$ and $g_s$, such that the $d$-wave gap magnitude satisfies
Eq.\ref{e7}, and the gap magnitude of extended-$s$-wave pairing channel vanishes at zero strain. Thus, in
this calculation, $d$-wave pairing dominates over the extended-$s$-wave pairing. The gap magnitude as a function
of uniaxial strain is shown in the left panel of Fig.\ref{f4}.  We solved the two gap equations, and
obtained the critical temperature and critical field.

$T_c$ and $H_{c2}$ as a function of uniaxial strain are summarized in the right panel of Fig.\ref{f4}.
Enhancement of $H_{c2}$ is notably stronger than of $T_c$. Further, under the
strain-independent BCS interaction, enhancement in $T_c$ and $H_c$ at the Van Hove strain agrees
quantitatively with experimental observations, with an maximal enhancement around 2.5 to 3 times. The
peak position matches with the Van Hove strain, at around $\tau=1.055$.

 It is worth noting that $g_s/g_d$ is a free parameter in the calculation. In the calculation of d+extended-s-wave pairing channel, we have chosen $g_d/g_s=1$, and the calculation of $d$-wave only pairing channel in Sec.\ref{Secd3} can be thought as special case with $g_s/g_d=0$. Choices of $g_s/g_d$ do not qualitatively change the results; in the calculation of stronger extended-s-wave pairing channel with $g_s/g_d=6.7$, where $g_s$ is taken such that the extended-$s$-wave gap magnitude satisfies
 Eq.\ref{e7}, $H_{c2}/T_c$ ratio also increases from zero strain to Van Hove strain by about $30\%$.

In this subsection, we have illustrated our numerical results on the ratio $H_{c2}/T_c$ as a function of
uniaxial strain, for different pairing channels. By comparing the results with experimental
observations, we found that $d$+extended-$s$-wave pairing channel can provide the correct
strain-dependence in the ratio $H_{c2}/T_c$ while $p$-wave pairing channel cannot.

Comparing the results in these two subsections, we summarize that the multi-band nature of
$\text{Sr}_2\text{Ru}\text{O}_4$ and $d$-wave type pairing channels are the key to explain the strain
dependence of the ratio $H_{c2}/T_c$.

\section{single-band FFLO state}\label{FFLO}
An inhomogeneous Fulde-Ferrell-Larkin-Ovchinnikov (FFLO) state, with non-zero momentum Cooper pairs, may
appear as an intermediate phase in strong applied fields under conditions that the Zeeman Effect
dominates over orbital suppression of the superconducting state \cite{Kun1998}. Such conditions otherwise apply in
very anisotropic organic superconductors for in-plane fields \cite{Wright2011}. And since the
applicability of an otherwise isotropic Zeeman effect implies singlet-pairing, the evidence for such an
intermediate state has been searched for in the case of parallel fields in \sro. In only one case that
we know of is there a suggestion for a field-induced intermediate phase \cite{Kikugawa2016}.
Nevertheless, we would like to study the possibility of the $d$-wave FFLO state on the $d_{xy}$ orbital
considered here.

Following \cite{Kun1998}, we extend the BCS interaction to
\begin{equation}
\begin{split}
H_{\text{BCS}}=-\sum_{\vec{k},\vec{k}',\vec{q},\lbrace{\sigma_i}\rbrace}V_{\sigma_1\sigma_2\sigma_3\sigma_4}(\vec{k},\vec{k}'){c}_{-\vec{k}\sigma_1}{c}_{\vec{k}+\vec{q}\sigma_2}{c}_{\vec{k}'+\vec{q}\sigma_3}^{\dagger}c_{-\vec{k}'\sigma_4}^{\dagger}
\end{split}
\end{equation}
The q-dependence in $V(\vec{k},\vec{k}')$ has been neglected. With the assumption of the separable
interaction in Eq.\ref{e5}, and $d$-wave pairing channel ${f_k}=i\sigma_y(\cos{k_x}-\cos{k_y})$, the BCS
gap equation can be simplified to
\begin{equation}
\begin{split}
1=&\sum_{|\epsilon_k|<E_{D}}g\frac{(\cos{k_x}-\cos{k_y})^2}{E_k}\\
&\left[\tanh{\frac{E_{k+q/2}+{H}}{2kT}}
+\tanh{\frac{E_{k-q/2}-{H}}{2kT}}\right]
\end{split}
\end{equation}
$H_{c2}$ for the FFLO state is the largest field among all possible $\vec{q}$, in which the gap equation has
solution $\Delta=0$. In this section, we choose interaction strength $g$ such that the zero momentum
state satisfies Eq.\ref{e7} with $E_D=50\Delta_0$.

At zero temperature and zero strain, we found that $H_{c2}\approx42\Delta_0$ at
$\vec{q}\approx{1.8H/v_{F,0}}\hat{x}(\text{or }\hat{y})$. At Van Hove strain, we found
$H_{c2}\approx39\Delta_0$ at $\vec{q}\approx{2.2H/v_{F,VH}}\hat{x}$. Here $v_{F,0}$ and $v_{F,VH}$ are
the Fermi velocity at $k_y=0$ for zero and Van Hove strain.

The magnitude of the FFLO critical field is found to be much higher than that of the zero-momentum
state. This may be due to Fermi surface nesting. The two ``vertical" parts in the $d_{xy}$ orbital (See
Fig.\ref{f1}) contributes to nesting, and therefore prefer a horizontal pairing momentum $\vec{q}$. This also
explains why the FFLO state is not sensitive to the Van Hove singularity, which is not part of the nesting.

Since the critical field of FFLO state is not sensitive to the Van Hove singularity, the $H_{c2}/T_c$ ratio for
FFLO state is therefore inconsistent with experimental observations for
$\text{Sr}_2\text{Ru}\text{O}_4$. However, our study of $H_{c2}$ as a function of uniaxial strain points out a new direction to
search for FFLO states in other materials.

\section{conclusion and discussion}
We studied the ratio $H_{c2}/T_c$ as a function of uniaxial strain for $\text{Sr}_2\text{Ru}\text{O}_4$,
and tried to match the experimentally observed increased ratio near the Van Hove strain.
We considered a three-band tight-binding Hamiltonian with separable and strain-independent BCS
interaction on the $d_{xy}$ orbital.
We tried different pairing channels and found that the experimental observation can be explained with
$d$+extended-$s$-wave, rather than $p$-wave pairing state.
We removed $d_{yz}$ and $d_{xz}$ orbitals, and then found that none of pairing channels could match the
experimental results.
Therefore, we concluded that the multi-band nature of $\text{Sr}_2\text{Ru}\text{O}_4$ and $d$-wave type
pairing channels are the key to explain the strain dependence of the ratio $H_{c2}/T_c$.

We also studied the ratio $H_{c2}/T_c$ for the FFLO state. Due to Fermi surface nesting, $d$-wave FFLO
state on $d_{xy}$ orbital is not sensitive to the Van Hove strain. However, our study points out a new way to
test FFLO state for a broader systems.

Lastly, we  discuss our results within the context of the broader phenomenological paradoxes presented by \sro.  Prior
to the NMR spectroscopy results in Ref. \onlinecite{Pustogow2019}, the key phenomenological issues involved rationalizing various experimental observations within the hypothesis of a
 chiral $p_x+ip_y$
superconducting ground state.  The experimental results in Ref. \onlinecite{Pustogow2019} have ruled out this scenario.  Instead, the focus has
shifted towards reconciling the NMR measurements with observations of TRSB in Kerr and muon spectroscopy studies.

On the one hand, TRSB requires having two distinct and degenerate order parameters.  This can be ensured by symmetry, if the order parameter belongs to a multi-dimensional irreducible representation (irrep).  Such states however, exhibit a split transition in the presence of the uniaxial strain considered in this paper: the absence of such split transitions casts significant doubt on the viability of such explanations.  TRSB can also occur in a fine-tuned situation where two distinct irreps become degenerate (see for instance the recent proposal in Ref. \onlinecite{2020arXiv200200016K}).    Such
degeneracy, if present, would be sensitive to perturbations, and may well be lifted by strain.  Indeed,  a recent mu-SR experiment in the presence of uniaxial strain shows the absence of TRSB at the superconducting transition from the normal state.\cite{grinenko2020split} It is thus
reasonable to start with a simpler setting of a single pairing channel when study the strain effects,
which is precisely what we have done here.

\section*{Acknowledgements}

KY's work was supported by the National Science Foundation Grant No. DMR-1932796. He also thanks the
National High Magnetic Field Laboratory, which is supported by National Science Foundation Cooperative
Agreement No. DMR-1644779, and the State of Florida. KY and SB also thank the Stanford Institute of Theoretical Physics and Gordon and Betty Moore Foundation for partial support. SB acknowledges partial support for this work from the National Science Foundation Grant No. DMR-1709304.  SR was  supported by the US
Department of Energy, Office of Basic Energy Sciences, Division of Materials Sciences and
Engineering, under contract number DE-AC02-76SF00515.

\bibliography{citation,SRO-Hc2-sb}{}

\begin{thebibliography}{21}
\expandafter\ifx\csname natexlab\endcsname\relax\def\natexlab#1{#1}\fi
\expandafter\ifx\csname bibnamefont\endcsname\relax
  \def\bibnamefont#1{#1}\fi
\expandafter\ifx\csname bibfnamefont\endcsname\relax
  \def\bibfnamefont#1{#1}\fi
\expandafter\ifx\csname citenamefont\endcsname\relax
  \def\citenamefont#1{#1}\fi
\expandafter\ifx\csname url\endcsname\relax
  \def\url#1{\texttt{#1}}\fi
\expandafter\ifx\csname urlprefix\endcsname\relax\def\urlprefix{URL }\fi
\providecommand{\bibinfo}[2]{#2}
\providecommand{\eprint}[2][]{\url{#2}}

\bibitem[{\citenamefont{Mackenzie and Maeno}(2003)}]{Mackenzie2003}
\bibinfo{author}{\bibfnamefont{A.~P.} \bibnamefont{Mackenzie}}
  \bibnamefont{and} \bibinfo{author}{\bibfnamefont{Y.}~\bibnamefont{Maeno}},
  \bibinfo{journal}{Rev. Mod. Phys.} \textbf{\bibinfo{volume}{75}},
  \bibinfo{pages}{657} (\bibinfo{year}{2003}).

\bibitem[{\citenamefont{Mackenzie et~al.}(2017)\citenamefont{Mackenzie,
  Scaffidi, Hicks, and Maeno}}]{Mackenzie2017}
\bibinfo{author}{\bibfnamefont{A.~P.} \bibnamefont{Mackenzie}},
  \bibinfo{author}{\bibfnamefont{T.}~\bibnamefont{Scaffidi}},
  \bibinfo{author}{\bibfnamefont{C.~W.} \bibnamefont{Hicks}}, \bibnamefont{and}
  \bibinfo{author}{\bibfnamefont{Y.}~\bibnamefont{Maeno}},
  \bibinfo{journal}{npj Quantum Materials} \textbf{\bibinfo{volume}{2}}
  (\bibinfo{year}{2017}).

\bibitem[{\citenamefont{Ishida et~al.}(1998)\citenamefont{Ishida, Mukuda,
  Kitaoka, Asayama, Mao, Mori, and Maeno}}]{Ishida1998}
\bibinfo{author}{\bibfnamefont{K.}~\bibnamefont{Ishida}},
  \bibinfo{author}{\bibfnamefont{H.}~\bibnamefont{Mukuda}},
  \bibinfo{author}{\bibfnamefont{Y.}~\bibnamefont{Kitaoka}},
  \bibinfo{author}{\bibfnamefont{K.}~\bibnamefont{Asayama}},
  \bibinfo{author}{\bibfnamefont{Z.~Q.} \bibnamefont{Mao}},
  \bibinfo{author}{\bibfnamefont{Y.}~\bibnamefont{Mori}}, \bibnamefont{and}
  \bibinfo{author}{\bibfnamefont{Y.}~\bibnamefont{Maeno}},
  \bibinfo{journal}{Nature} \textbf{\bibinfo{volume}{396}},
  \bibinfo{pages}{658} (\bibinfo{year}{1998}).

\bibitem[{\citenamefont{Duffy et~al.}(2000)\citenamefont{Duffy, Hayden, Maeno,
  Mao, Kulda, and McIntyre}}]{Duffy2000}
\bibinfo{author}{\bibfnamefont{J.~A.} \bibnamefont{Duffy}},
  \bibinfo{author}{\bibfnamefont{S.~M.} \bibnamefont{Hayden}},
  \bibinfo{author}{\bibfnamefont{Y.}~\bibnamefont{Maeno}},
  \bibinfo{author}{\bibfnamefont{Z.}~\bibnamefont{Mao}},
  \bibinfo{author}{\bibfnamefont{J.}~\bibnamefont{Kulda}}, \bibnamefont{and}
  \bibinfo{author}{\bibfnamefont{G.~J.} \bibnamefont{McIntyre}},
  \bibinfo{journal}{Phys. Rev. Lett.} \textbf{\bibinfo{volume}{85}},
  \bibinfo{pages}{5412} (\bibinfo{year}{2000}).

\bibitem[{\citenamefont{Luke et~al.}(1998)\citenamefont{Luke, Fudamoto, Kojima,
  Larkin, Merrin, Nachumi, Uemura, Maeno, Mao, Mori et~al.}}]{Luke1998}
\bibinfo{author}{\bibfnamefont{G.~M.} \bibnamefont{Luke}},
  \bibinfo{author}{\bibfnamefont{Y.}~\bibnamefont{Fudamoto}},
  \bibinfo{author}{\bibfnamefont{K.~M.} \bibnamefont{Kojima}},
  \bibinfo{author}{\bibfnamefont{M.~I.} \bibnamefont{Larkin}},
  \bibinfo{author}{\bibfnamefont{J.}~\bibnamefont{Merrin}},
  \bibinfo{author}{\bibfnamefont{B.}~\bibnamefont{Nachumi}},
  \bibinfo{author}{\bibfnamefont{Y.~J.} \bibnamefont{Uemura}},
  \bibinfo{author}{\bibfnamefont{Y.}~\bibnamefont{Maeno}},
  \bibinfo{author}{\bibfnamefont{Z.~Q.} \bibnamefont{Mao}},
  \bibinfo{author}{\bibfnamefont{Y.}~\bibnamefont{Mori}}, \bibnamefont{et~al.},
  \bibinfo{journal}{Nature} \textbf{\bibinfo{volume}{394}},
  \bibinfo{pages}{558} (\bibinfo{year}{1998}).

\bibitem[{\citenamefont{Xia et~al.}(2006)\citenamefont{Xia, Maeno, Beyersdorf,
  Fejer, and Kapitulnik}}]{Xia2006}
\bibinfo{author}{\bibfnamefont{J.}~\bibnamefont{Xia}},
  \bibinfo{author}{\bibfnamefont{Y.}~\bibnamefont{Maeno}},
  \bibinfo{author}{\bibfnamefont{P.~T.} \bibnamefont{Beyersdorf}},
  \bibinfo{author}{\bibfnamefont{M.~M.} \bibnamefont{Fejer}}, \bibnamefont{and}
  \bibinfo{author}{\bibfnamefont{A.}~\bibnamefont{Kapitulnik}},
  \bibinfo{journal}{Phys. Rev. Lett.} \textbf{\bibinfo{volume}{97}}
  (\bibinfo{year}{2006}).

\bibitem[{\citenamefont{Pustogow et~al.}(2019)\citenamefont{Pustogow, Luo,
  Chronister, Su, Sokolov, Jerzembeck, Mackenzie, Hicks, Kikugawa, Raghu
  et~al.}}]{Pustogow2019}
\bibinfo{author}{\bibfnamefont{A.}~\bibnamefont{Pustogow}},
  \bibinfo{author}{\bibfnamefont{Y.}~\bibnamefont{Luo}},
  \bibinfo{author}{\bibfnamefont{A.}~\bibnamefont{Chronister}},
  \bibinfo{author}{\bibfnamefont{Y.-S.} \bibnamefont{Su}},
  \bibinfo{author}{\bibfnamefont{D.~A.} \bibnamefont{Sokolov}},
  \bibinfo{author}{\bibfnamefont{F.}~\bibnamefont{Jerzembeck}},
  \bibinfo{author}{\bibfnamefont{A.~P.} \bibnamefont{Mackenzie}},
  \bibinfo{author}{\bibfnamefont{C.~W.} \bibnamefont{Hicks}},
  \bibinfo{author}{\bibfnamefont{N.}~\bibnamefont{Kikugawa}},
  \bibinfo{author}{\bibfnamefont{S.}~\bibnamefont{Raghu}},
  \bibnamefont{et~al.}, \bibinfo{journal}{Nature}
  \textbf{\bibinfo{volume}{574}}, \bibinfo{pages}{72} (\bibinfo{year}{2019}).

\bibitem[{\citenamefont{Ishida et~al.}(2020)\citenamefont{Ishida, Manago,
  Kinjo, and Maeno}}]{Ishida2020}
\bibinfo{author}{\bibfnamefont{K.}~\bibnamefont{Ishida}},
  \bibinfo{author}{\bibfnamefont{M.}~\bibnamefont{Manago}},
  \bibinfo{author}{\bibfnamefont{K.}~\bibnamefont{Kinjo}}, \bibnamefont{and}
  \bibinfo{author}{\bibfnamefont{Y.}~\bibnamefont{Maeno}},
  \bibinfo{journal}{Journal of the Physical Society of Japan}
  \textbf{\bibinfo{volume}{89}}, \bibinfo{pages}{034712}
  (\bibinfo{year}{2020}).

\bibitem[{\citenamefont{Petsch et~al.}(2020)\citenamefont{Petsch, Zhu, Enderle,
  Mao, Maeno, and Hayden}}]{Petsch2020}
\bibinfo{author}{\bibfnamefont{A.~N.} \bibnamefont{Petsch}},
  \bibinfo{author}{\bibfnamefont{M.}~\bibnamefont{Zhu}},
  \bibinfo{author}{\bibfnamefont{M.}~\bibnamefont{Enderle}},
  \bibinfo{author}{\bibfnamefont{Z.~Q.} \bibnamefont{Mao}},
  \bibinfo{author}{\bibfnamefont{Y.}~\bibnamefont{Maeno}}, \bibnamefont{and}
  \bibinfo{author}{\bibfnamefont{S.~M.} \bibnamefont{Hayden}},
  \emph{\bibinfo{title}{Reduction of the spin susceptibility in the
  superconducting state of \sro\ observed by polarized neutron scattering}}
  (\bibinfo{year}{2020}), \eprint{2002.02856}.

\bibitem[{\citenamefont{Hicks et~al.}(2014)\citenamefont{Hicks, Brodsky,
  Yelland, Gibbs, Bruin, Barber, Edkins, Nishimura, Yonezawa, Maeno
  et~al.}}]{Hicks2014}
\bibinfo{author}{\bibfnamefont{C.~W.} \bibnamefont{Hicks}},
  \bibinfo{author}{\bibfnamefont{D.~O.} \bibnamefont{Brodsky}},
  \bibinfo{author}{\bibfnamefont{E.~A.} \bibnamefont{Yelland}},
  \bibinfo{author}{\bibfnamefont{A.~S.} \bibnamefont{Gibbs}},
  \bibinfo{author}{\bibfnamefont{J.~A.~N.} \bibnamefont{Bruin}},
  \bibinfo{author}{\bibfnamefont{M.~E.} \bibnamefont{Barber}},
  \bibinfo{author}{\bibfnamefont{S.~D.} \bibnamefont{Edkins}},
  \bibinfo{author}{\bibfnamefont{K.}~\bibnamefont{Nishimura}},
  \bibinfo{author}{\bibfnamefont{S.}~\bibnamefont{Yonezawa}},
  \bibinfo{author}{\bibfnamefont{Y.}~\bibnamefont{Maeno}},
  \bibnamefont{et~al.}, \bibinfo{journal}{Science}
  \textbf{\bibinfo{volume}{344}}, \bibinfo{pages}{283} (\bibinfo{year}{2014}).

\bibitem[{\citenamefont{Steppke et~al.}(2017)\citenamefont{Steppke, Zhao,
  Barber, Scaffidi, Jerzembeck, Rosner, Gibbs, Maeno, Simon, Mackenzie
  et~al.}}]{Steppke2017}
\bibinfo{author}{\bibfnamefont{A.}~\bibnamefont{Steppke}},
  \bibinfo{author}{\bibfnamefont{L.}~\bibnamefont{Zhao}},
  \bibinfo{author}{\bibfnamefont{M.~E.} \bibnamefont{Barber}},
  \bibinfo{author}{\bibfnamefont{T.}~\bibnamefont{Scaffidi}},
  \bibinfo{author}{\bibfnamefont{F.}~\bibnamefont{Jerzembeck}},
  \bibinfo{author}{\bibfnamefont{H.}~\bibnamefont{Rosner}},
  \bibinfo{author}{\bibfnamefont{A.~S.} \bibnamefont{Gibbs}},
  \bibinfo{author}{\bibfnamefont{Y.}~\bibnamefont{Maeno}},
  \bibinfo{author}{\bibfnamefont{S.~H.} \bibnamefont{Simon}},
  \bibinfo{author}{\bibfnamefont{A.~P.} \bibnamefont{Mackenzie}},
  \bibnamefont{et~al.}, \bibinfo{journal}{Science}
  \textbf{\bibinfo{volume}{355}} (\bibinfo{year}{2017}), ISSN
  \bibinfo{issn}{0036-8075},
  \eprint{https://science.sciencemag.org/content/355/6321/eaaf9398.full.pdf},
  \urlprefix\url{https://science.sciencemag.org/content/355/6321/eaaf9398}.

\bibitem[{\citenamefont{Li et~al.}(2019)\citenamefont{Li, Kikugawa, Sokolov,
  Jerzembeck, Gibbs, Maeno, Hicks, Nicklas, and Mackenzie}}]{Li2019}
\bibinfo{author}{\bibfnamefont{Y.~S.} \bibnamefont{Li}},
  \bibinfo{author}{\bibfnamefont{N.}~\bibnamefont{Kikugawa}},
  \bibinfo{author}{\bibfnamefont{D.~A.} \bibnamefont{Sokolov}},
  \bibinfo{author}{\bibfnamefont{F.}~\bibnamefont{Jerzembeck}},
  \bibinfo{author}{\bibfnamefont{A.~S.} \bibnamefont{Gibbs}},
  \bibinfo{author}{\bibfnamefont{Y.}~\bibnamefont{Maeno}},
  \bibinfo{author}{\bibfnamefont{C.~W.} \bibnamefont{Hicks}},
  \bibinfo{author}{\bibfnamefont{M.}~\bibnamefont{Nicklas}}, \bibnamefont{and}
  \bibinfo{author}{\bibfnamefont{A.~P.} \bibnamefont{Mackenzie}},
  \emph{\bibinfo{title}{High precision heat capacity measurements on \sro\
  under uniaxial pressure}} (\bibinfo{year}{2019}), \eprint{1906.07597}.

\bibitem[{\citenamefont{Grinenko et~al.}(2020)\citenamefont{Grinenko, Ghosh,
  Sarkar, Orain, Nikitin, Elender, Das, Guguchia, Br{\"u}ckner, Barber
  et~al.}}]{grinenko2020split}
\bibinfo{author}{\bibfnamefont{V.}~\bibnamefont{Grinenko}},
  \bibinfo{author}{\bibfnamefont{S.}~\bibnamefont{Ghosh}},
  \bibinfo{author}{\bibfnamefont{R.}~\bibnamefont{Sarkar}},
  \bibinfo{author}{\bibfnamefont{J.-C.} \bibnamefont{Orain}},
  \bibinfo{author}{\bibfnamefont{A.}~\bibnamefont{Nikitin}},
  \bibinfo{author}{\bibfnamefont{M.}~\bibnamefont{Elender}},
  \bibinfo{author}{\bibfnamefont{D.}~\bibnamefont{Das}},
  \bibinfo{author}{\bibfnamefont{Z.}~\bibnamefont{Guguchia}},
  \bibinfo{author}{\bibfnamefont{F.}~\bibnamefont{Br{\"u}ckner}},
  \bibinfo{author}{\bibfnamefont{M.~E.} \bibnamefont{Barber}},
  \bibnamefont{et~al.}, \bibinfo{journal}{arXiv preprint arXiv:2001.08152}
  (\bibinfo{year}{2020}).

\bibitem[{\citenamefont{Luo et~al.}(2019)\citenamefont{Luo, Pustogow, Guzman,
  Dioguardi, Thomas, Ronning, Kikugawa, Sokolov, Jerzembeck, Mackenzie
  et~al.}}]{Luo2019}
\bibinfo{author}{\bibfnamefont{Y.}~\bibnamefont{Luo}},
  \bibinfo{author}{\bibfnamefont{A.}~\bibnamefont{Pustogow}},
  \bibinfo{author}{\bibfnamefont{P.}~\bibnamefont{Guzman}},
  \bibinfo{author}{\bibfnamefont{A.~P.} \bibnamefont{Dioguardi}},
  \bibinfo{author}{\bibfnamefont{S.~M.} \bibnamefont{Thomas}},
  \bibinfo{author}{\bibfnamefont{F.}~\bibnamefont{Ronning}},
  \bibinfo{author}{\bibfnamefont{N.}~\bibnamefont{Kikugawa}},
  \bibinfo{author}{\bibfnamefont{D.~A.} \bibnamefont{Sokolov}},
  \bibinfo{author}{\bibfnamefont{F.}~\bibnamefont{Jerzembeck}},
  \bibinfo{author}{\bibfnamefont{A.~P.} \bibnamefont{Mackenzie}},
  \bibnamefont{et~al.}, \bibinfo{journal}{Phys. Rev. X}
  \textbf{\bibinfo{volume}{9}}, \bibinfo{pages}{021044} (\bibinfo{year}{2019}),
  \urlprefix\url{https://link.aps.org/doi/10.1103/PhysRevX.9.021044}.

\bibitem[{\citenamefont{Yonezawa et~al.}(2014)\citenamefont{Yonezawa, Kajikawa,
  and Maeno}}]{Yonezawa2014}
\bibinfo{author}{\bibfnamefont{S.}~\bibnamefont{Yonezawa}},
  \bibinfo{author}{\bibfnamefont{T.}~\bibnamefont{Kajikawa}}, \bibnamefont{and}
  \bibinfo{author}{\bibfnamefont{Y.}~\bibnamefont{Maeno}}, \bibinfo{journal}{J.
  Phys. Soc. Jpn.} \textbf{\bibinfo{volume}{83}}, \bibinfo{pages}{083706}
  (\bibinfo{year}{2014}).

\bibitem[{\citenamefont{Zabolotnyy et~al.}(2013)\citenamefont{Zabolotnyy,
  Evtushinsky, Kordyuk, Kim, Carleschi, Doyle, Fittipaldi, Cuoco, Vecchione,
  and Borisenko}}]{Zabolotnyy2013}
\bibinfo{author}{\bibfnamefont{V.}~\bibnamefont{Zabolotnyy}},
  \bibinfo{author}{\bibfnamefont{D.}~\bibnamefont{Evtushinsky}},
  \bibinfo{author}{\bibfnamefont{A.}~\bibnamefont{Kordyuk}},
  \bibinfo{author}{\bibfnamefont{T.}~\bibnamefont{Kim}},
  \bibinfo{author}{\bibfnamefont{E.}~\bibnamefont{Carleschi}},
  \bibinfo{author}{\bibfnamefont{B.}~\bibnamefont{Doyle}},
  \bibinfo{author}{\bibfnamefont{R.}~\bibnamefont{Fittipaldi}},
  \bibinfo{author}{\bibfnamefont{M.}~\bibnamefont{Cuoco}},
  \bibinfo{author}{\bibfnamefont{A.}~\bibnamefont{Vecchione}},
  \bibnamefont{and}
  \bibinfo{author}{\bibfnamefont{S.}~\bibnamefont{Borisenko}},
  \bibinfo{journal}{Journal of Electron Spectroscopy and Related Phenomena}
  \textbf{\bibinfo{volume}{191}}, \bibinfo{pages}{48 } (\bibinfo{year}{2013}),
  ISSN \bibinfo{issn}{0368-2048},
  \urlprefix\url{http://www.sciencedirect.com/science/article/pii/S0368204813001655}.

\bibitem[{\citenamefont{Chandrasekhar}(1962)}]{chandrasekhar1962}
\bibinfo{author}{\bibfnamefont{B.}~\bibnamefont{Chandrasekhar}},
  \bibinfo{journal}{Applied Physics Letters} \textbf{\bibinfo{volume}{1}},
  \bibinfo{pages}{7} (\bibinfo{year}{1962}).

\bibitem[{\citenamefont{Yang and Sondhi}(1998)}]{Kun1998}
\bibinfo{author}{\bibfnamefont{K.}~\bibnamefont{Yang}} \bibnamefont{and}
  \bibinfo{author}{\bibfnamefont{S.~L.} \bibnamefont{Sondhi}},
  \bibinfo{journal}{Phys. Rev. B} \textbf{\bibinfo{volume}{57}},
  \bibinfo{pages}{8566} (\bibinfo{year}{1998}),
  \urlprefix\url{https://link.aps.org/doi/10.1103/PhysRevB.57.8566}.

\bibitem[{\citenamefont{Wright et~al.}(2011)\citenamefont{Wright, Green, Kuhns,
  Reyes, Brooks, Schlueter, Kato, Yamamoto, Kobayashi, and Brown}}]{Wright2011}
\bibinfo{author}{\bibfnamefont{J.~A.} \bibnamefont{Wright}},
  \bibinfo{author}{\bibfnamefont{E.}~\bibnamefont{Green}},
  \bibinfo{author}{\bibfnamefont{P.}~\bibnamefont{Kuhns}},
  \bibinfo{author}{\bibfnamefont{A.}~\bibnamefont{Reyes}},
  \bibinfo{author}{\bibfnamefont{J.}~\bibnamefont{Brooks}},
  \bibinfo{author}{\bibfnamefont{J.}~\bibnamefont{Schlueter}},
  \bibinfo{author}{\bibfnamefont{R.}~\bibnamefont{Kato}},
  \bibinfo{author}{\bibfnamefont{H.}~\bibnamefont{Yamamoto}},
  \bibinfo{author}{\bibfnamefont{M.}~\bibnamefont{Kobayashi}},
  \bibnamefont{and} \bibinfo{author}{\bibfnamefont{S.~E.} \bibnamefont{Brown}},
  \bibinfo{journal}{Phys. Rev. Lett.} \textbf{\bibinfo{volume}{107}}
  (\bibinfo{year}{2011}).

\bibitem[{\citenamefont{Kikugawa et~al.}(2016)\citenamefont{Kikugawa,
  Terashima, Uji, Sugii, Maeno, Graf, Baumbach, and Brooks}}]{Kikugawa2016}
\bibinfo{author}{\bibfnamefont{N.}~\bibnamefont{Kikugawa}},
  \bibinfo{author}{\bibfnamefont{T.}~\bibnamefont{Terashima}},
  \bibinfo{author}{\bibfnamefont{S.}~\bibnamefont{Uji}},
  \bibinfo{author}{\bibfnamefont{K.}~\bibnamefont{Sugii}},
  \bibinfo{author}{\bibfnamefont{Y.}~\bibnamefont{Maeno}},
  \bibinfo{author}{\bibfnamefont{D.}~\bibnamefont{Graf}},
  \bibinfo{author}{\bibfnamefont{R.}~\bibnamefont{Baumbach}}, \bibnamefont{and}
  \bibinfo{author}{\bibfnamefont{J.}~\bibnamefont{Brooks}},
  \bibinfo{journal}{Phys. Rev. B} \textbf{\bibinfo{volume}{93}},
  \bibinfo{pages}{184513} (\bibinfo{year}{2016}).

\bibitem[{\citenamefont{{Kivelson} et~al.}(2020)\citenamefont{{Kivelson},
  {Yuan}, {Ramshaw}, and {Thomale}}}]{2020arXiv200200016K}
\bibinfo{author}{\bibfnamefont{S.~A.} \bibnamefont{{Kivelson}}},
  \bibinfo{author}{\bibfnamefont{A.~C.} \bibnamefont{{Yuan}}},
  \bibinfo{author}{\bibfnamefont{B.~J.} \bibnamefont{{Ramshaw}}},
  \bibnamefont{and}
  \bibinfo{author}{\bibfnamefont{R.}~\bibnamefont{{Thomale}}},
  \bibinfo{journal}{arXiv e-prints} \bibinfo{eid}{arXiv:2002.00016}
  (\bibinfo{year}{2020}), \eprint{2002.00016}.

\end{thebibliography}
\end{document}